\documentstyle[12pt]{article}
\begin{document}
\title{
A possible phase diagram of a  t-J ladder model }
\author{ 
Kazuhiro SANO   \\
 Department of Physics, Faculty of Education, Mie University, \\
  Tsu, Mie 514.      }

\date{(Received \hspace{6cm})}

\maketitle


\begin{abstract}
\baselineskip=20pt

We investigate   a t-J ladder model by numerical diagonalization method. By
calculating correlation functions and assuming the  Luttinger liquid
relation, we obtained a possible phase diagram of the ground state as a
function of J/t and electron density $n$. 
 We also found that behavior  of correlation functions seems to consist
with the prediction of Luttinger liquid relation.
 The result suggests that the superconducting phase appear in the region of
$J/t  \displaystyle{ \mathop{>}_{\sim}} 0.5$ for high electron density  and
 $J/t  \displaystyle{ \mathop{>}_{\sim}} 2.0$ for low electron density.
 
\end{abstract}


\newpage

 Low-dimensional antiferomagnetic  systems    attract much interest due to
the possible relevance to high-$T_c$ superconductivity.
In particular, ladder  systems which show a finite spin gap are extensively
studied now. 
For example, recent measurements of the magnetic susceptibility and/or the
nuclear relaxation rate of SrCu$_2$O$_3$ and (VO)$_2$P$_2$O$_7$ indicate
that these materials are described  weakly coupled ladder  models with
spin-gap\cite{MTakano,Ishida,Barnes}. 
 Many  theoretical works also   have been performed on   Heisenberg of
ladder models and the spin gap up to order of exchange energy is
found\cite{Dagotto,Gopalan,Troyer1}.
  Recently, It is  claimed that the spin-gap survives against  light  hole
doping and it vanishes for heavy doping  in  $t$-$J$ ladder models by
numerical diagonalization method.\cite{Tsunetsugu,Poilblanc,Troyer2}
Hayward et al. gave numerical evidence of  superconductivity in  $t$-$J$
ladders for  sufficiently large exchange energy $J$ by a density matrix
renormalization method\cite{Hayward1}.

 If  the low energy behavior of the ladder system  can be  described as
that of one-dimensional (1D) system, an approach of the 'Luttinger liquid'
theory  will be useful.
In the Luttinger liquid theory,  some relations  have been  established as
universal relations \cite{Haldane,Haldane2}. 
They supplied us with unambiguous and important information about not only
one-band electronic systems but  complicated 2-band  systems\cite{Sano}.

 Hayward and Poilblanc \cite{Hayward2} discussed the Luttinger liquid
behavior of a t-J ladder model and give a possible phase diagram of  a
critical exponent  $K_\rho$ which characterizes the algebraic decay  of
correlation functions in the  Luttinger liquid theory.
 However, it is not clear whether or not  exponents of correlation
functions given by the Luttinger liquid relation consist with the behavior
of  correlation functions calculated by the numerical
method\cite{Hayward1}.

 In this letter,  we reexamine the critical exponent   and  Luttinger
liquid behavior of t-J ladder   by   numerical diagonalization method.  
We find that a prediction of the Luttinger liquid theory consists with  the
behavior of the correlation functions of t-J ladder.  
Using the exponent,  we determine  a superconducting region on the 
possible phase diagram. 

The t-J ladder Hamiltonian  is written as 
\begin{eqnarray*}
H=-t\sum_{i}( c^{\dagger}_{i,\alpha,\sigma} c_{i+1,\alpha,\sigma} + h.c.)
-t\sum_{i}( c^{\dagger}_{i,1,\sigma} c_{i,2,\sigma} + h.c.) \\
+ J\sum_{i,\alpha}({\bf S}_{i,\alpha}\cdot {\bf S}_{i+1,\alpha}
-\frac{1}{4}n_{i,\alpha} n_{i,\alpha})
+J\sum_{i}( {\bf S}_{i,1}\cdot {\bf S}_{i,2} -\frac{1}{4}n_{i,1} n_{i,2} )
\hspace{2mm}, \quad \quad \quad (1)
\end{eqnarray*}  
 where $c^{\dagger}_{i\sigma}$ is the electron creation operator
   with spin $\sigma$ on a rung $i$ of ladder
and ${\bf S}_{i,\alpha}$ is the spin operator made of
$c^{\dagger}_{i,\alpha,\sigma}$ and $c_{i,\sigma,\alpha}$. Here,
$\alpha(=1,2)$ labels the two legs  of the ladder and 
$n_{i,\alpha}=\sum_{\sigma} c^{\dagger}_{i,\alpha,\sigma} c_{i,\alpha,\sigma}$. 
 $J$ is exchange energy between the nearest neighbor sites and  $t$  stands
for the transfer energy, which will be set to be unity  hereafter in the
present study.  
We take account of the infinite on-site repulsion by removing states with
doubly occupied sites from the Hilbert space. 
 
 We numerically diagonalize the Hamiltonian  up to 20 sites (10 unit 
cells)  using the standard Lanczos algorithm. 
  We use the periodic boundary  condition for $N_e=4m+2$ and anti periodic
boundary condition for $N_e=4m$, where $N_e$ is the  total number of
electrons and m is  an  integer. 
    We also use the open boundary  condition to calculate  correlation
functions because we are able to take more points  in  correlation
functions.
    The filling $n$ is defined  by  $n=N_{e}/(2N_{u})$, where  $N_u$ is the
total number of unit cells ( each unit cell corresponds to a rung ).
 
   The chemical potential $\mu (N_{e},N_{u})$ is defined by
$$
   \mu (N_{e},N_{u})=\frac{E_{0}(N_{e}+1,N_u)-E_{0}(N_{e}-1,N_u)}2\quad ,
\eqno{(2)}
$$
where $E_{0}(N_e,N_u)$ is the total ground state energy.
     When the charge gap vanishes in the thermodynamic limit, the uniform 
charge susceptibility $\chi_c$ is obtained from 
$$
 \chi_c(N_{e},N_{u})=\frac{2/N_u}{\mu(N_{e}+1,N_{u})-\mu(N_{e}-1,N_{u})} 
\quad .\eqno{(3)}
$$
It is noted that our definition of the $\chi_c$ is larger than that of
previous works\cite{Troyer2,Hayward2} by a factor of $2$ since we  use the
charge susceptibility $\chi_c$ per rung rather than per site.

 In  the model which is isotropic in spin space,   exponents of various
types of correlation functions are determined by the critical exponent
$K_\rho$.
For single band model, the Luttinger liquid theory predicts that
Superconducting(SC) correlation function is dominant for $K_{\rho}>1$
(attractive case), whereas Charge or Spin Density Wave (CDW or SDW)
correlation is dominant  for $K_{\rho}<1$ (repulsive case).
In the case of non-interacting fermion systems, the exponent $K_{\rho}$ is
always unity.

On the other hand, the bosonization method\cite{Schulz,Balents} shows that
two-chain model with a small interchain hopping has a spin-gaped phase and
SC  and $"4k_F"$ CDW correlations decay  as $\sim r^{-\frac{1}{2K_{\rho}}}$
and $\sim r^{-2K_{\rho}}$ respectively (SDW and $"2k_F"$ CDW correlations
decay exponentially).  
Hence,  SC correlation is dominant for  $K_\rho >0.5$. 

When we apply the above relations to the t-J ladder, we should pay
attention to the filling of electrons in the t-J ladder. 
For low density region, we should adopt the  relation for the single band
model since electrons  fill in only the lower band of the t-J ladder. 
For high density region, we can expect the  relation for two-chain model to
be useful.   

 It is convenient to introduce a critical exponent
  $\tilde{K}_{\rho}$ which is related to the usual critical exponent 
$K_{\rho}$ by the relations ,
$$
          \tilde{K}\sb{\rho}=\frac{\pi}{2}v\sb{c}\chi\sb{c} \quad ,  \eqno{(4)}
$$
$$
           v_c=\frac{N_u}{2\pi}(E_{1}-E_{0}) \quad ,  \eqno{(5)}
$$
where $E_{1}-E_{0}$ is the lowest charge excitation energy.
For low density, our definition of the $\tilde{K}_\rho$ is equivalent  to
that of the  exponent $K_\rho$ \cite{Haldane,Ogata}. 
When electrons are filled in only  lower band,  we should define the
$K_\rho$ of  a multi-band model as above.
Another definition of $K_\rho$ in previous works \cite{Troyer2,Hayward2} 
is misleading in applying the Luttinger liquid relation to t-J ladder at
low density. 
On the contrary, for high density, the $\tilde{K}_\rho$ is larger than the
$K_\rho$  by a factor of $2$ since we use the $\chi_c$ given by eq.(3).
Considering  the $\chi_c$ and $K_{\rho}$ of a non-interacting  ladder
system, it is easily understood. 

In both the low density  and the high density cases, we expect that the
superconducting correlations  dominate for $\tilde{K}_\rho > 1$.

We can also determine the $\tilde{K}_{\rho}$ by the Drude weight $D$
$$           
      \tilde{K}\sb{\rho}=\pi(\frac{\chi_c D}{2})^{1/2} \quad ,    \eqno{(6)}
$$
$$
D=\frac{1}{2N_u} \frac{\partial^2 E(\phi)}{\partial \phi^2}.\quad \eqno{(7)}
$$
where $E(\phi)$ is the total energy of the ground state as a function of flux 
$\phi$.
Using these two independent equations of the $\tilde{K}_{\rho}$, we can 
check the consistency of the Luttinger liquid relations.\cite{Hayward2+}   

To investigate the band structure of  the t-J ladder model, we show the
chemical potential $\mu$ ( Fermi energy $E_F$ at the $T=0$ ) as a function
of electron filling $n$ in Fig.1. 
For a non-interacting ladder model, band structure is   given as 
  $$ E^{\pm}(k)=  2tcos(k)  \pm t    \eqno{(8)}$$  
   where $k$ is a wave vector. 
When $n$ is smaller than 0.5, electrons are filled in only lower band $
E^{-}(k)$. 
At $n=0.5$, lower and upper bands  begin to fill    with electrons
simultaneously.
The slope of $\mu$   is equal to zero and the value of the charge
susceptibility $\chi_c$ diverges at $n=0.5$, where the slope of $\mu$
corresponds to  $2/\chi_c$.
Thus, the divergence  of $\chi_c$  signifies a change in the electronic
state of non-interacting  systems.

On the other hand, $\chi_c$  of t-J ladders seems to be finite at $n=0.5$
for $J \displaystyle{ \mathop{<}_{\sim}}1.0$.
If the divergent point  of $\chi_c$ exists, it may move to more high density. 
In  contrast to non-interacting fermion ladder, we expect that  the ground
state of t-J ladder  continues from low density over $n=0.5$. 
In fact,  points of $n=0.5$ for $J \displaystyle{ \mathop{<}_{\sim}}1.0$
belongs to the spin-gapless region, which is yield by Ref.
\cite{Poilblanc}.
Roughly speaking, Ref.\cite{Poilblanc} shows that  the spin-gap phase
appears  at high density and large $J$  and the remainder  is the gapless
phase.

It is interesting to see that the $n-$dependence of $\mu$ at $J/t=2$ 
almost corresponds to that of non-interacting  ladder system  in low
electron density.  
It suggests that the electronic state of non-interacting   system 
resembles that of t-J ladder in this parameter region.  To confirm the
above, we calculate the density-density correlation function on the same
leg $C_{CDW}(R)=<n_{0,\alpha} n_{R,\alpha}>-<n_{0,\alpha}>< n_{R,\alpha}>$
of both systems, where  $R$ is distance from one of  ends of the system.  
In Fig.2(a), we show $C_{CDW}(R)$ for 18-site systems with 4-electrons
under the open boundary condition.
Correlation functions of both systems are very close to each other. We also
get similar results for singlet pairing and spin-spin correlation functions
of the same systems. 

These results show that low energy behavior of t-J ladder is approximately
described as non-interacting system. It also indicates that the correlation
exponent $\tilde{K}_\rho (=K_\rho)$  is close to unity.
The direct numerical calculation of $\tilde{K}_\rho$ by eq.(4)  shows 
$\tilde{K}_\rho \simeq 1.0$ for $N_s=18$ system with $N_e$=4.
It  reminds us that the wave function of a single chain t-J model at
$t=J/2$ is close to projected  Fermi liquid\cite{Yokoyama} and the exponent
$K_\rho$  becomes unity in the low  density limit\cite{Ogata}. 
 Therefore, the properties of t-J ladder are close to that of the  single
band  model.
 It suggests that the existence  of upper band $ E^{+}(k)$ is irrelevant to
the low energy behavior of t-J ladder in the low electron density
region\cite{Balents}.

Next, we consider the case of high electron density. 
For  $J=1$, we can also  compare a behavior of correlation functions and
the value of $\tilde{K}_\rho(=2K_\rho)$. 
Recently, Hayward et al. calculated   correlation functions by using the
density-matrix renormalization-group method at $J/t=1$ and $n=0.8$. Their
results show that a pairing correlation is longer range than  other
correlations,  decaying slower  than $\sim R^{-1}$.
The power  of the density-density correlation function seems to be $-2$ or
slightly larger. 

If we assume  $\tilde{K}_\rho \sim 1.5$ in this case, the powers of 
correlation functions can be explained with the 'duality' relation which is
introduced by Nagaosa and Oshikawa\cite{Nagaosa}; the powers of paring and
density-density correlations are obtained as $\sim 0.7$ and  as $\sim 1.5$
respectively.
It is consist with  the value of  $\tilde{K}_\rho \sim 1.4$, which is
obtained from the Luttinger liquid relation  for the $N_s=20$ site system
with 16 electrons. 
These results suggest that the Luttinger liquid relation holds  for t-J ladder. 

 Furthermore, we calculate the pairing correlations and  density-density
correlations for another system.
Figs.3 show the rung-rung pairing correlation function $C_{SC}(R)=<\Delta_0
\Delta^\dagger_R>$, where $\Delta^\dagger_R= (c^{\dagger}_{R,1,\uparrow}
c^{\dagger}_{R,2,\downarrow}- c^{\dagger}_{R,1,\downarrow}
c^{\dagger}_{R,2,\uparrow})$ and the density-density correlation function
$C_{CDW}(R)$ for $N_s=16$ site system at $n$=12/16.
For $J=0.5$, both correlation functions seem to decay as $\sim R^{-1}$ as
shown in Fig.3(a). 
For $J=1.5$, pairing correlations decay  slower than  the density-density
correlations; the former seems to decay as $\sim R^{-0.5}$ or more rapidly
and the later seems to decay as $\sim R^{-2}$ as shown in Fig.3(b).

 Although the system size is not sufficiently large  to get  exponents of
correlation functions precisely, the powers of both correlation functions
seem to satisfy the duality relation.
At least, the behavior of correlation functions suggests that the paring
correlations dominate others for  $J/t \displaystyle{
\mathop{>}_{\sim}}0.5$.
Using eq.(6), the critical exponent $\tilde{K}_\rho$ is obtained as
$\sim1.0$ and $\sim2.0$ for $J=0.5$ and $J=1.5$ respectively.  It consists
with the  behavior of correlation functions and indicates  the validity of
the Luttinger liquid relation for two-chain model.

In the limit of $J\to 0$, the situation is complicated.
 Nagaoka ferromagnetic state is known to appear at the ground state of
finite systems\cite{Troyer2}. 
However, a very small $J$  lifts the   ferromagnetic state and  changes the
ground state to singlet.
If the  Nagaoka ferromagnetic phase exists in the thermodynamic limit, the
phase may be restricted to the region with very small $J$. 
Thus,  the lowest singlet state is not the true ground state at $J=0$;
nevertheless,  we single it out as the relevant state at very small $J$.

Fig.1 shows that the band structure of t-J ladder at $J=0$ is close to that
of non-interacting spinless fermions in the ladder model. 
In Fig.4, we compare the density-density correlation functions $C_{CDW}(R)$
of t-J ladder with that of spinless fermion ladder. 
It shows that both correlation functions are close to each other.
Using the equations (6), the correlation exponent $\tilde{K}_\rho$  is
estimated as $\sim0.4$\cite{polynomial}, which is almost consist with  the
value of $K_\rho (=0.5)$ of the spinless fermion system.
 The same situation is already found in a single chain t-J model for  very
small $J$\cite{Ogata-Shiba}. 
This result indicates that we should apply not the relation for two-chain
model but single-chain model  contrary to expectation.
Probably, only the upper band $ E^{+}(k)$ is relevant to the low energy
behavior  and   charge degree of freedom is reduced to that of  spinless
fermion system  in the high  density region at $J=0$.

Finally, we examine the phase diagram of superconducting state in the
$J$-$n$ plane  with the exponent $\tilde{K}_\rho$. 
In Fig.5, we show the possible phase diagram of t-J ladder. When the
$\tilde{K}_\rho$ ($\chi_c$) diverges, the uniform state becomes  unstable
and phase separation occurs.
The  boundary of phase separation  agrees with the results of previous
works\cite{Tsunetsugu,Troyer2}.
Fig.5 shows that the region of superconducting state appears at lower value
of $J/t$ than a single chain case at high electron density.
 It seems to resemble the behavior in two-dimensional case. 
On the other hand, the behavior of the superconducting region at small
electron concentration resembles that of a single chain\cite{Ogata}.
Hayward et al.\cite{Hayward2} also obtained the phase diagram in the
$J$-$n$ plane, but  their value of $K_\rho$ is smaller than
$\tilde{K}_\rho$  by a factor 2.
If the value of $K_\rho$ is doubled in their phase diagram, the result of
$\tilde{K}_\rho$  agrees with ours.  

The author gratefully acknowledges many useful discussions with Takami
Tohyama. The computation in this work has been done using the facilities of
the Supercomputer Center, Institute for Solid State Physics, University of
Tokyo.

\newpage
\noindent
Figure captions

\bigskip

Fig.1. The chemical potential $\mu$ as a function of the filling $n$ for
various values of $J$. 
 $\mu$  is calculated for 
$N_s$=  12, 14, 16, 18,and 20. 
The  dashed line represents a  non-interacting fermion ladder  band and the
broken line represents a spinless-fermion ladder  band.

\bigskip

Fig.2. The density-density correlation functions $C_{CDW}(R)$ of t-J ladder
and non-interacting fermion  systems.
 $C_{CDW}(R)$  is calculated for the $N_s$= 18 sites system with 4 electrons. 
The dashed  line has  slope -2.

\bigskip
Fig.3.
The rang-rang paring correlation function $C_{SC}(R)$ and the
density-density correlation function $C_{CDW}(R)$ of the $N_s$= 16 sites
system with 12 electrons (a) for $J/t=0.5$  and (b) for $J/t=1.5$.

\bigskip
Fig.4. 
 The density-density correlation functions $C(R)_{CDW}$ of t-J ladder and
non-interacting spinless-fermion  systems.
 $C(R)_{CDW}$ is calculated for the $N_s$= 18 sites system with 16 electrons. 
The dashed  line has a slope -1.5.

\bigskip

Fig.5. A possible phase diagram of the t-J ladder as a function of $J/t$
and electron density $n=N_e/N_s$.
The circles  represent results of  $N_s$= 16 sites systems. 


\begin{thebibliography}{99}
\bibitem{Hayward2+} 
Detailed analysis for t-J ladder is shown in Ref.\cite{Hayward2}.
\bibitem{polynomial} 
In the high density region, it is difficult to obtain $\chi_c$ precisely by
using eq. (4) since frustration of  antiferomagnetic  configuration on the
legs increases. Thus, we fit data of $\mu$  to a polynomial  as a function
of $n$ by the least square method and estimate the $\chi_c$ from
differential coefficient of the polynomial. Using the above $\chi_c$, we
calculated  the $\tilde{K}_\rho$ by the eq.(6).
\end{thebibliography}
\end{document}